\documentclass[accepted]{uai2025} 
                        
\usepackage[american]{babel}

\usepackage{natbib} 
\usepackage{mathtools} 
\usepackage{booktabs} 

\usepackage{tikz} 
\usetikzlibrary{arrows,shapes,snakes,automata,backgrounds,petri,arrows.meta}
\tikzset{>={Classical TikZ Rightarrow[length=1mm]}}

\usepackage{subcaption}
\DeclareCaptionLabelFormat{sublabel}{\text{#2)}\hspace{1.5ex}}

\newcommand{\mmax}{m_\text{max}}
\newcommand{\mest}{m_\text{est}}
\newcommand{\mtrue}{m_\text{true}}
\newcommand{\TP}{\textit{TP}}
\newcommand{\FP}{\textit{FP}}
\newcommand{\TN}{\textit{TN}}
\newcommand{\FN}{\textit{FN}}
\newcommand{\tp}{\textit{tp}}
\newcommand{\fp}{\textit{fp}}
\newcommand{\tn}{\textit{tn}}
\newcommand{\fn}{\textit{fn}}

\newcommand{\E}{\text{E}}

\DeclareMathOperator{\cond}{|}

\title{Are You Doing Better Than Random Guessing? A Call for Using Negative Controls When Evaluating Causal Discovery Algorithms}

\author[1]{\href{mailto:<ahpe@sund.ku.dk>}{Anne~Helby~Petersen}}
\affil[1]{%
    Section of Biostatistics\\
    Department of Public Health\\
    University of Copenhagen\\
    Copenhagen, Denmark
}

\begin{document}
\maketitle

\begin{abstract}
New proposals for causal discovery algorithms are typically evaluated using simulations and a few selected real data examples with known data generating mechanisms. However, there does not exist a general guideline for how such evaluation studies should be designed, and therefore, comparing results across different studies can be difficult. In this article, we propose to use negative controls as a common evaluation baseline by posing the question: Are we doing better than random guessing? For the task of graph skeleton estimation, we derive exact distributional results under random guessing for the expected behavior of a range of typical causal discovery evaluation metrics, including precision and recall. We show that these metrics can achieve very favorable values under random guessing in certain scenarios, and hence warn against using them without also reporting negative control results, i.e., performance under random guessing. We also propose an exact test of overall skeleton fit, and showcase its use on a real data application. Finally, we propose a general pipeline for using negative controls beyond the skeleton estimation task, and apply it both in a simulated example and a real data application. 
\end{abstract}

\section{Introduction}
\label{sec.intro}
Causal discovery algorithms seek to infer information about a causal data generating mechanism by analyzing empirical data it generated. The causal data generating mechanism is typically represented by a causal graph, for example an equivalence class of directed acyclic graphs (DAGs). A highly productive research community has published a plethora of new causal discovery algorithms within the last 30 years or so. Naturally, this fast growing battery of available algorithms requires some standards and guidelines for evaluating and benchmarking their performance. Because the result of a causal discovery algorithm is an estimated graph (or family of graphs), rather than one or more scalars, it is not entirely obvious how to use classic approaches for performance evaluation from neither machine learning nor statistics. 

Nonetheless, machine learning classification metrics originally developed for evaluating prediction tasks are often used to evaluate causal discovery algorithms. Most commonly, precision and recall, or possibly their harmonic mean, the F1 score, are reported, although some studies also focus on other metrics, e.g., negative predictive value \citep{petersen2023sldisco}. These metrics are computed from graph-level confusion matrices summarizing either agreement on placement of oriented edges (primarily used for DAG discovery evaluation), adjacencies (i.e., edge placement without considering orientation), and/or arrowheads among correctly placed adjacencies (conditional orientation). Typically, they are reported as averages over numerous simulations. Sometimes the results are stratified by graphical parameters (e.g., true graph density), data-related parameters (e.g., sample size), or simply reported as averages across several such settings. 

Alternative metrics developed specifically for graphs also exist; the structural Hamming distance \citep{tsamardinos2006} is the most widely used example, probably due to its cheap computation and easy interpretation. A different metric focusing more on the causal implications of the graphs is the structural intervention distance (SID) \citep{peters2015}, although it is most naturally suited for DAG-DAG comparison, and hence not readily applicable for all discovery evaluation tasks. A more recent proposal is the adjustment identification distance \citep{henckel2024}, which also focuses on differences in causal inference based on the graph, or the separation distance \citep{wahl2025}, which counts agreement in separation statements. There are thus many different possible choices of metrics for evaluating causal discovery algorithms. However, no general guidelines exist on how to then \textit{interpret} the values these metrics take: What is a high or low number? 

An often-used strategy for answering this question in experimental sciences is to conduct a controlled experiment. In such an experiment, the intervention of interest (here: a causal discovery algorithm) is compared to a control condition. When this control condition cannot have any influence on the outcome of interest, it is denoted a \textit{negative control}. For example, say we want to study the impact of a fertilizer on plant growth. We plant 100 seeds in two plots with similar conditions, except that one (the treatment group) receives the fertilizer, while the other (the negative control) does not. After, say, 10 days, we measure the heights of the plants, and we use the average difference in heights as a measure of the effect of the fertilizer. By including the negative control we obtain a direct measure of the specific effect of the treatment.

Alternatively, we could have compared the fertilizer of interest with another active treatment, perhaps an alternative well-known fertilizer (denoted a \textit{positive control}). This comparison gives less information about the specific treatment of interest. For example, we cannot falsify a hypothesis saying that neither fertilizer has any effect. In some scientific fields, for example human drug trials, using positive controls is the only viable option, as it would be utterly unethical to deny patients treatment (if one exists) in order to obtain a negative control group for evaluating a new proposed treatment. But of course, no such concerns are relevant for evaluating causal discovery algorithms. Nonetheless, the current standard practice is to report positively controlled experiments: A new candidate algorithm is typically compared to a selection of existing algorithms. Such a comparison in itself does not provide information about whether \textit{either} of the considered algorithms work. Moreover, because causal discovery evaluations are very sensitive towards experimental settings concerning graph sparsity (as we will demonstrate below), it is not straight-forward to generalize findings from such a positively controlled experiment to infer what performance should be expected on just slightly different evaluation settings. This makes it very difficult to compare results across different evaluation studies with just marginally different designs.

We propose to use negative controls to obtain an interpretable benchmark for any causal discovery evaluation study: Namely, to investigate what values of the metrics of interest can be obtained using random guessing (a negative control), and report this alongside findings from positive controls (alternative algorithms). Others have reported results from a single random guess alongside causal discovery benchmarks \citep{lachapelle2020}, but as we will argue, a more structured approach to including negative controls provides many benefits. We discuss negative controls in two different settings: First, we consider the task of skeleton estimation, that is estimating e.g., a DAG without taking orientation information into account. For this case, we derive exact distributional results for the expected behavior under random guessing (Section \ref{sec.distrresults}), and we use these results to compute expected negative control values for a range of often-used metrics (Section \ref{subsec.expectations-exact}). We furthermore propose an exact test of overall skeleton fit (Section \ref{subsec.testfit}), and provide an example of its use on real data. Secondly, we consider more general metrics that are not only concerned with skeleton estimation, and propose a negative control pipeline for this case (Section \ref{sec.simbasedNC}). We provide two examples of its use, both in a simulation study and in a real data application. 

But before we turn to these general results, we present an example case where well-known metrics such as adjacency precision and recall do perhaps not behave exactly as one would have expected. 

Code for all computations is available online at \url{https://github.com/annennenne/negcontrol-disco}. 

\section{Precision and Recall: A Cautionary Tale}
\label{sec.precrecall}

\begin{figure}[t]
\centering
\begin{subfigure}[b]{0.22\textwidth}
\centering
\caption{True DAG}\label{fig.d5ex.a}
\begin{tikzpicture}[scale = 0.7]
[every edge/.append style={nodes={font=\tiny}}]
\node (1) at (3,5) {$X_1$};
\node (2) at (1,3) {$X_2$};
\node (3) at (2,1) {$X_3$};
\node (4) at (4,1) {$X_4$};
\node (5) at (5,3) {$X_5$};
\draw [->] (1) edge (2);
\draw [->] (1) edge (4);
\draw [->] (1) edge (5);
\draw [->] (2) edge (3);
\draw [->] (2) edge (4);
\draw [->] (2) edge (5);
\draw [->] (4) edge (5);
\draw [->] (5) edge (3);
\end{tikzpicture}
\end{subfigure}
\begin{subfigure}[b]{0.22\textwidth}
\centering
\caption{Estimated DAG}\label{fig.d5ex.b}
\begin{tikzpicture}[scale = 0.7]
[every edge/.append style={nodes={font=\tiny}}]
\node (1) at (3,5) {$X_1$};
\node (2) at (1,3) {$X_2$};
\node (3) at (2,1) {$X_3$};
\node (4) at (4,1) {$X_4$};
\node (5) at (5,3) {$X_5$};
\draw [->] (1) edge (2);
\draw [->] (1) edge (3);
\draw [->] (1) edge (4);
\draw [->] (1) edge (5);
\draw [->] (3) edge (2);
\draw [->] (4) edge (2);
\draw [->] (5) edge (3);
\end{tikzpicture}
\end{subfigure}
\caption{The True Underlying DAG (a) and an Estimated DAG (b) Obtained Using an Undisclosed Causal Discovery Procedure.}\label{fig.d5ex}
\end{figure}

\begin{table}
\caption{Adjacency Confusion Matrix for the 5 Node DAG Example in Figure \ref{fig.d5ex}.}
\label{tab.conf5node}
\begin{center}
\small
\begin{tabular}{llcc}
& & \multicolumn{2}{c}{\textbf{Truth}} \\
&  & Adjacency & Non-adjacency \\
\hline
\textbf{Estimate} & Adjacency & $\tp = 6$ & $\fp = 1$ \\
& Non-adjacency & $\fn = 2$ & $\tn = 1$ \\
\hline
\end{tabular}
\end{center}
\end{table}

Consider the two DAGs in Figure \ref{fig.d5ex}. The left graph (a) is the true DAG, and the right graph (b) is an estimate produced by a causal discovery procedure. We compute their adjacency confusion matrix in order to evaluate the performance of the discovery procedure (Table \ref{tab.conf5node}). This results in: 
\begin{align*}
\text{precision} &= \frac{\tp}{\tp + \fp} = \frac{6}{7} \simeq 0.86  \quad \quad &\text{ and } \\
\quad  \quad  \text{recall} &=  \frac{\tp}{\tp + \fn} = \frac{6}{8} = 0.75.
\end{align*}
Are these numbers high or low? Although these values are not too far off from the performance of well-established causal discovery algorithms on simulated data (and much better than typical performance on "real" benchmarking datasets), we will argue that they are indeed as low as can be for this specific discovery task --- because the "discovery algorithm" applied here was simply random guessing and hence had absolutely no information about the true data generating mechanism.

More specifically, we simulated 1000 random Erdős-Rényi type DAGs over 5 nodes each with 7 edges\footnote{Note that this is not even the correct number of edges, although close to it, as the true DAG has 8 edges.} and used these "random guesses" as estimates of the DAG in Figure \ref{fig.d5ex} (a). This resulted in a median precision of 0.86 and a median recall of 0.75, i.e., numbers that exactly match the performance of the example just described. The DAG shown in Figure 1 (b) was one among many random draws that matches this median performance. Hence the large values of precision and recall cannot be attributed to a conveniently chosen random seed. 

Is it then a curious artefact for very dense graphs? Or "small" graphs over e.g., 5 nodes? Neither is the case. As we will show in the following section, the phenomenon does not depend on the number of nodes, and depending on the choice of metrics, can occur also in modestly dense graphs.

\section{Distributional Results for Adjacency Metrics Under Random Guessing}
\label{sec.distrresults}

\begin{table}
\small
\centering
\caption{Generically Labelled Adjacency Confusion Matrix.}
\label{tab.genericconf}
\begin{tabular}{llccc}
& & \multicolumn{3}{c}{\textbf{Truth}} \\
&  & Adjacency & Non-adjacency & Total\\
\hline
\textbf{Est.} & Adjacency & $\TP$ & $\FP$ & $\mest$ \\
& Non-adjacency & $\FN$ & $\TN$ & - \\
& Total & $\mtrue$ & - & $\mmax$ \\
\hline
\end{tabular}
\caption*{\smaller Notes: \textit{Entries marked with dashes are sums that will not be used for the derivations here. "Est." abbreviates estimate.}}
\end{table}

Consider a DAG $G$ over $d$ nodes, and let $\mtrue$ denote the number of edges in $G$. Let $\hat{G}$ be another DAG over the same $d$ nodes used as an estimate of $G$, and let $\mest$ be the number of edges in $\hat{G}$. Finally, let $\mmax = \sum_{i = 1}^{d-1} i = \frac{1}{2} (d - 1) d$ denote the maximal number of possible edges in a DAG over $d$ nodes (corresponding to a fully connected graph). We can describe the performance of $\hat{G}$ as an adjacency/skeleton estimator of $G$ through a (generic) confusion matrix as seen in Table \ref{tab.genericconf}. Note that for a given causal discovery problem, namely estimating some given $G$, $\mmax$ and $\mtrue$ can be considered fixed: $\mmax$ depends only on the number of nodes $d$, which does not change, and $\mtrue$ is fixed given $G$. Moreover, for many causal discovery procedures, it further makes sense to consider $\mest$ fixed --- at least for a specific value of a tuning parameter (e.g., significance level for testing or penalty for a score) and a specific dataset --- as we most often do not try to \textit{estimate} the correct number of edges from data. Rather, algorithms are typically applied with a pre-specified value of the tuning parameter (e.g. significance level of $\alpha = 0.01$), which indirectly controls the resulting number of edges, $\mest$ (see Sections \ref{sec.simbasedNC} and \ref{sec.disc} for considerations in cases where this latter assumption is not meaningful). 

We now make the following important observation: If edges are placed uniformly in both $G$ and  $\hat{G}$ (corresponding to Erdős-Rényi type graphs), and we condition on the row and column sums of Table \ref{tab.genericconf} (which is equivalent with conditioning on $\mtrue$, $\mest$ and $\mmax$), then by definition, the number of true positives will follow a hypergeometric distribution parameterized by $m_\text{max}$,  $m_\text{true}$ and $m_\text{est}$: 
\begin{equation*} 
\TP  \cond \mmax, \mtrue, \mest \sim \text{HyperGeom}(\mmax, \mtrue, \mest).
\end{equation*}
Note that this is an exact distributional result, not an asymptotic statement. This distributional result is well known from its use in the (one-sided) Fisher's exact test, and may also be motivated using a random urn experiment analogy, see Supplementary Materials \ref{suppl.randomurn}. 

This observation gives rise to several useful applications: First, we can compute the expected value, median and uncertainty estimates (e.g., confidence interval) for the number of true positive adjacencies under random guessing. Secondly, since we are also conditioning on $m_\text{max}$, $m_\text{true}$ and $\mest$, we can further compute expectations and draw statistical inference for any function of the confusion matrix, including precision, recall and F1. We provide formulas for these in Section \ref{subsec.expectations-exact}. Thirdly, we can construct an exact statistical test of overall skeleton fit by considering how much the number of true positives in a given estimated graph diverts from its expected distribution under a null hypothesis of random edge placement. We propose such a test in Section \ref{subsec.testfit}. 

\section{Expectations and Quantiles of Adjacency Metrics Under Random Guessing}
\label{subsec.expectations-exact}

\begin{table}
\begin{center}
\caption{Expected Values and Quantile Expressions Under Random Guessing for Five Commonly Used Adjacency Metrics.}
\label{tab.adjmetrics}
\begin{tabular}{lcc} 
\hline 
\textbf{Metric} & \textbf{Expected value} & \textbf{Quantile}  \\
 \hline  
 Precision & $\frac{\mtrue}{\mmax}$ & $\frac{q_{k}}{\mest}$ \medskip \\
Recall & $\frac{\mest}{\mmax}$ & $\frac{q_{k}}{\mtrue}$ \medskip \\ 
F1 & $\frac{2 \cdot \mest \cdot \mtrue}{\mmax \cdot \mest + \mmax \cdot \mtrue}$ & $\frac{ 2 \cdot q_{k}}{\mest + \mtrue}$  \medskip \\
NPV &  $1 - \frac{\mtrue}{\mmax}$ & $\frac{\mmax - \mest - \mtrue + q_{k}}{\mmax - \mest}$  \medskip \\ 
Specificity &  $1 - \frac{\mest}{\mmax}$ & $\frac{\mmax - \mest - \mtrue + q_{k}}{\mmax - \mtrue}$  \medskip \\ 
\hline 
\end{tabular}\end{center}
\caption*{\smaller Note: \textit{$q_{k}$ denotes the $k$th quantile from $\text{HyperGeom}(\mmax, \mtrue, \mest)$.}}
\end{table}

Since 
\begin{equation*}
\TP \cond \mmax, \mtrue, \mest \sim \text{HyperGeom}(\mmax, \mtrue, \mest),
\end{equation*}
by definition we have that 
\begin{equation*}
\E(\TP  \cond \mmax, \mtrue, \mest) =  \frac{\mest \cdot \mtrue}{\mmax}
\end{equation*}
and by considering the quantile function of $\text{HyperGeom}(\mmax, \mtrue, \mest)$, we can construct a confidence interval as e.g., the central 95\% of the distribution, or find the expected median. 

For fixed values of $(\mmax, \mtrue, \mest)$, Table \ref{tab.adjmetrics} provides an overview of expected values and quantiles under random guessing for five metrics commonly used for evaluating adjacency placement for causal discovery algorithms, namely precision, recall, F1 score, negative predictive value (NPV) and specificity. As an example, we here showcase derivations for precision, and refer to Supplementary Materials \ref{suppl.adjmetrics} for derivations for the remaining four metrics.

\paragraph{Expectation and quantiles for adjacency precision}
We first express precision as a function of $\TP, \mmax, \mtrue$ and $\mest$: 
\begin{align*}
\text{prec} &= \frac{\TP}{\TP + \FP} = \frac{\TP}{\TP + \mest - \TP} = \frac{\TP}{\mest}
\end{align*}
Since this is a linear function of $\TP$, we can straight-forwardly compute the expectation: 
\begin{align*}
\E(\text{prec} \cond \mmax, \mtrue, \mest) &= \frac{1}{\mest} \E(\TP \cond \mmax, \mtrue, \mest) \\
&= \frac{\mtrue}{\mmax}
\end{align*}
The linearity also makes it easy to obtain e.g., an exact 95\% confidence interval under the null hypothesis of random guessing by simply applying the same transformation to the appropriate quantiles of $\text{HyperGeom}(\mmax, \mtrue, \mest)$. For example, an exact 95\% confidence interval for precision under the null is given by
\begin{equation*}
 \left( \frac{1}{\mest} q_{(0.025,\mmax, \mtrue, \mest)},\frac{1}{\mest} q_{(0.975,\mmax, \mtrue, \mest)}\right)
\end{equation*}
where $q_{(k,\mmax, \mtrue, \mest)}$ is used to denote the $k$th quantile of the probability mass function of $\text{HyperGeom}(\mmax, \mtrue, \mest)$. Similarly, we obtain the median precision by simply computing
\begin{equation*}
\text{median}(\text{prec}) = \frac{1}{\mest} q_{(0.5,\mmax, \mtrue, \mest)}.
\end{equation*}

\paragraph{General remarks concerning Table \ref{tab.adjmetrics}}
A notable feature of Table \ref{tab.adjmetrics} is that, conditional on $(\mmax, \mtrue, \mest)$, the expected precision is simply the density of the true DAG $G$, and the expected recall is the density of the estimated DAG $\hat{G}$. Furthermore, the expected values of NPV and specificity are given as 1 minus the expectations of precision and recall, respectively, and hence they do not provide additional information. However, without random guessing, this is of course not generally the case, so they are still useful to compute in order to provide a nuanced and multifaceted evaluation of a given causal discovery procedure. 

Moreover, we note that under random guessing, the expected precision does not depend on the number of edges in the estimated graph ($\mest$), only on the number of edges in the true graph ($\mtrue$) and the maximal possible number of edges ($\mmax$). But recall increases linearly as a function of the number of estimated edges. Hence, if we are using random guessing, a "free lunch" in optimizing precision and recall is achievable simply by estimating a very large number of edges, even including the trivial fully connected graph. This can also be seen from the expected value of the F1 score under random guessing, which increases monotonically with the number of estimated edges: It is always better to just add another edge.

Note that all statistical inference based on Table \ref{tab.adjmetrics} relies on a single distributional result. Hence, while we may construct confidence intervals for e.g., both precision and recall to aid our own interpretation of a specific estimation problem, there is no additional statistical information gained: We might as well just draw inference directly on the number of true positives. 

We will now consider two small example applications of the results from Table \ref{tab.adjmetrics}. First, we revisit the example from Section \ref{sec.precrecall} and compute the median, expected value, and a 95\% confidence interval for precision and recall for this case. Next, we provide an overview of how the expected F1 score varies as a function of $\mest$ and $\mtrue$ under random guessing. 

\subsection{Example: Expected Adjacency Precision and Recall for a Dense 5 Node DAG}
Consider the problem from Section \ref{sec.precrecall} regarding estimating a DAG skeleton over 5 nodes. Such a DAG can have at most $\mmax = \frac{1}{2} (5-1) 5 = 10$ edges. Assume that the true DAG has $\mtrue = 8$ edges, while a randomly drawn graph over the same 5 nodes has $\mest = 7$ edges. What performance can we then expect from this random guessing procedure? With reference to Table \ref{tab.adjmetrics}, we find 
\begin{equation*}
\E(\text{prec} \cond \mmax = 10, \mtrue = 8, \mest = 7) =  \frac{8}{10} = 0.80
\end{equation*}
with a 95\% confidence interval of 
\begin{equation*}
 \left( \frac{q_{(0.025,10, 8, 7)}}{7} ,\frac{q_{(0.975,10,8,7)}}{7}  \right) =  \left( \frac{5}{7},\frac{7}{7}  \right) \simeq (0.71, 1.00).
\end{equation*}
Hence for this DAG estimation task, it will not be highly unusual to obtain adjacency precisions as high as 1.00 under random guessing, and thus adjacency precision is not very useful for assessing performance. We can also compute the median precision:
\begin{equation*}
\text{median}(\text{prec}) =  \frac{q_{(0.5,10, 8, 7)}}{7}  = \frac{6}{7} \simeq 0.86
\end{equation*}
This is the same value as found in the simulations presented in Section \ref{sec.precrecall}.

For adjacency recall we find 
\begin{equation*}
\E(\text{recall} \cond \mmax = 10, \mtrue = 8, \mest = 7) =  \frac{7}{10} = 0.70
\end{equation*}
and we compute a 95\% confidence interval as 
\begin{equation*}
\left(\frac{q_{0.025,10, 8, 7}}{8} ,\frac{q_{0.975,10, 8, 7}}{8}  \right)  =   \left(  \frac{5}{8},\frac{7}{8} \right) \simeq (0.63, 0.88).
\end{equation*}
One is not included in this confidence interval and hence adjacency recall does have some discriminatory power for this DAG estimation task. We compute the median: 
\begin{equation*}
\text{median}(\text{recall}) = \frac{q_{(0.5,10, 8, 7)}}{8}  = \frac{6}{8} = 0.75
\end{equation*}
Once again, this matches our simulation-based findings from Section \ref{sec.precrecall}. 

\subsection{Example: Adjacency F1 Scores for a 5 Node DAG With Varying Density}

\begin{figure}[t]
\centering
\includegraphics[scale=0.5]{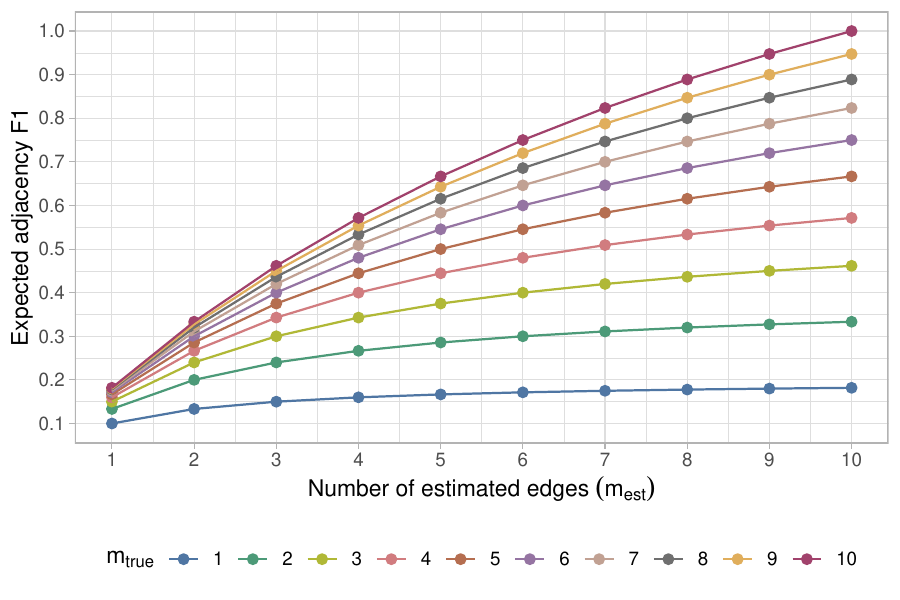}
\caption{Expected Adjacency F1 Scores Under Random Guessing for Estimating 5 Node DAGs.}
\label{fig.d5anres-f1}
\end{figure}

Figure \ref{fig.d5anres-f1} provides an overview of obtained F1 scores under random guessing across all possible combinations of estimated number of edges (horizontal axis) and true number of edges (marked in color) for 5 node DAGs. We see that it is quite possible to obtain a large F1 score by random guessing if the true DAG is not very sparse, and especially, if the estimate is also not very sparse. But we stress that neither has to be overly or unrealistically dense either: For a true graph that has just 5 edges --- i.e., only one edge more than the sparsest graph that is connected --- a randomly drawn DAG with 5 edges will result in an expected F1 of 0.5, and placing all 10 possible edges results in an F1 score of 0.66. If we instead consider a more dense graph, e.g., a true DAG with 8 edges, we are back in the scenario already considered above, and we see that we can find a peak F1 score of 0.89 by placing all edges.

\section{A Test of Overall Skeleton Fit}
\label{subsec.testfit}

We can also use the distributional results presented above to construct an exact test of overall skeleton fit. More specifically, for an estimated DAG $\hat{G}$ with $\mest$ edges, we test the null hypothesis
\begin{equation*}
H_0: \hat{G} \text{ was obtained by randomly placing } \mest \text{ edges.}
\end{equation*}
This is done by comparing the observed number of true positives, $\tp_\text{obs}$ with the appropriate hypergeometric distribution. Formally, if we let \mbox{$X \sim \text{HyperGeom}(\mmax, \mtrue, \mest)$}, a one-sided p-value for $H_0$ is computed as 
\begin{equation*}
P(X \geq \tp_\text{obs})
\end{equation*}
i.e., the probability of getting at least as many true positives as the observed number if edges were in fact randomly placed. Note that since the test is exact (and based on a discrete probability distribution), it will be conservative.

\subsection{Application: Temporal PC on the Metropolit Cohort Dataset}
\label{sec.application-metropolit}

\begin{table}
\centering
\caption{Adjacency Confusion Matrix Replicated From \cite{petersen2023}.}
\label{tab.metro}
\begin{tabular}{llcc}
& & \multicolumn{2}{c}{\textbf{Experts}} \\
&  & Adjacency & Non-adjacency \\
\hline
\textbf{TPC} & Adjacency & 10 & 20 \\
& Non-adjacency & 20 & 181 \\
\hline
\end{tabular}
\end{table}

We will reanalyze data from \cite{petersen2023}. In that study, the temporal PC algorithm (TPC) \citep{petersen2021} was used on a cohort data set of $n = 3145$ Danish men to identify possible causes of depression and heart disease, as well as their interplay. Two experts were also asked to construct a model for the data based on existing studies and subject-field knowledge, and their DAG was compared to the output of TPC. For the comparison here, we assume that the expert model is correct and wish to evaluate if TPC performs better than a negative control at estimating the expert model. The DAGs have 22 nodes, and hence $\mmax = 231$ possible edges. 

Table \ref{tab.metro} shows the adjacency confusion matrix comparing the expert and TPC models. Note that the two models did not disagree on edge orientation among shared adjacencies (although one shared adjacency was left unoriented by TPC). Hence in this case, the adjacency performance comparison summarizes all edge-wise comparisons of the two outputs. Note also that TPC was set to find the same number of edges as the experts did, i.e. $\mtrue = \mest = 30$, and hence the symmetry in the confusion matrix is by design. 

We conduct an overall test of skeleton fit by comparing the obtained number of true positives, $\tp_\text{obs} = 10$, with $\text{HyperGeom}(231, 30, 30)$ and we find $p = 0.002$. Hence, we reject $H_0$ and conclude that TPC performs significantly better than random guessing in this application.

\section{Simulation-Based Negative Controls for More General Metrics}
\label{sec.simbasedNC}

Although the results provided above cover some of the most commonly reported metrics for causal discovery evaluation, other interesting metrics cannot be expressed as functions of the adjacency confusion matrix, and hence are out the scope of the results presented thus far. 

One example is conditional orientation metrics (also sometimes referred to as \textit{arrowhead} metrics) (see e.g., \citet{andrews2019}). These metrics describe correct orientations among correctly placed edges. We conjecture that simple exact distributional results under random guessing do not exist for this classification task. The main issue is that consecutive edge placement steps are \textit{not} independent when the goal is to output e.g., a valid DAG: If we have already placed oriented edges such that $X \to Y \to Z$, it is no longer possible to have an edge pointing from $Z$ to $X$, as this would introduce a cycle and the graph would then no longer be a valid DAG. Thus, describing expected behavior under random guessing when also taking edge orientations into account is more complicated. 

However, we can easily use simulation to obtain an empirical estimate of the distribution of a given metric under random edge placement --- oriented or not. Let $b$ be the number of repetitions in the simulation study, and let $f$ denote some metric of interest.We propose the following procedure: 
\begin{description}
\item[1. Standard simulation study:]  Conduct the simulation study as usual: Simulate $b$ "true" DAGs $G_\text{true}^1, ..., G_\text{true}^b$, generate appropriate data for each, and use the causal discovery algorithm of interest to obtain estimated graphs $\hat{G}_\text{algo}^1, ..., \hat{G}_\text{algo}^b$. For each $i \in \{1, ..., b\}$, compare the true and estimated graphs by computing the metric of interest, $f(G_\text{true}^i, \hat{G}_\text{algo}^i)$, and for each estimated graph $\hat{G}_\text{algo}^i$, count the number edges, $\mest^i$. 
\item[2. Negative control simulation:] For each $i \in \{1, ..., b\}$, draw a negative control random DAG $\hat{G}_\text{NC}^i$ with number of edges sampled randomly from $\{\mest^1, ..., \mest^b\}$ (with replacement). 
\item[3. Negative control evaluation:] Compare each negative control with the corresponding true graph by computing the metric of interest, $f(G_\text{true}^i, \hat{G}_\text{NC}^i)$. Report the mean as the expected performance under random guessing, and use the empirical quantiles to construct e.g., a 95\% confidence interval.
\item[4. Comparison:] Finally, compare the metrics obtained under random guessing with the metrics obtained for the evaluated algorithm. In order to draw statistical inference, consider pairwise comparisons and conduct a one-sided statistical test. Compute the $p$-value as 
\begin{equation*}
p = \frac{1}{b}\sum_{i = 1}^b \mathbf{1} \left(f(G_\text{true}^i, \hat{G}_\text{algo}^i)  \leq f(G_\text{true}^i, \hat{G}_\text{NC}^i)\right)
\end{equation*}
for metrics where small values are favorable (otherwise reverse the inequality). 
\end{description}
Note that it is important for obtaining valid statistical inference that it is conducted on the pairwise comparisons of performance, as inference e.g., based on whether or not confidence intervals overlap is highly conservative \citep{knol2011}.

If the evaluated algorithm does not estimate a DAG, we suggest that Step 2 is altered to match the output of the evaluated algorithm. For example, if the algorithm only aims to learn the Markov equivalence class of the data generating DAG, as represented by a CPDAG, we would simply use negative control CPDAGs in Step 2 by first drawing DAGs and subsequently finding their encompassing CPDAGs.

We have here focused on the case of a simulation study where many different ground truth graphs are simulated in Step 1, but in Section \ref{subsec.sachs} we also provide an example of how to adapt the procedure to be suited for evaluation of a real data application where there is only a single ground truth.

\subsection{Example: Simulation Study Evaluating the PC Algorithm}
\begin{table*}
\centering
\caption{Comparisons of PC Algorithm and Negative Controls.}
\label{tab.pcsim}
\begin{tabular}{lrcrcc}
  \hline
 & \multicolumn{2}{c}{\textbf{PC}} & \multicolumn{2}{c}{\textbf{Negative control}} &  \\
 & Mean & CI & Mean & CI &  $p$\\ 
  \hline
\textbf{Dense case} ($\mtrue = 30$) \\  
 \quad SHD & 27.33 & $(21,33)$ & 31.23 & $(26,36)$ & 0.202 \\ 
 \quad Adjacency precision &  0.85 & $(0.65,1.00)$ &  0.66 & $(0.42,0.87)$ & 0.122 \\ 
 \quad Adjacency recall &  0.38 & $(0.27,0.50)$ &  0.29 & $(0.17,0.43)$ & 0.245 \\ 
 \quad Orientation precision &  0.65 & $(0,1)$ &  0.50 & $(0,1)$ & 0.360 \\ 
 \quad Orientation recall &  0.40 & $(0.00,0.78)$ &  0.37 & $(0.00,0.78)$ & 0.464 \\ 
 \quad Proportion recovered v-structures &  0.05 & $(0.0,0.2)$ &  0.02 & $(0.00,0.14)$ & 0.563 \\ 
 \quad SID (lower bound) & 67.73 & $(46,83)$ & 74.23 & $(56,85)$ & 0.317 \\ 
 \quad SID (upper bound) & 79.48 & $(61,90)$ & 79.10 & $(63,88)$ & 0.557 \\\hline 
\textbf{Sparse case} ($\mtrue = 15$) \\  
\quad SHD & 10.1 & $( 4,15)$ & 21.30 & $(17,25)$ & 0.002 \\ 
 \quad Adjacency precision &  0.9 & $(0.73,1.00)$ &  0.33 & $(0.091,0.571)$ & 0.000 \\ 
 \quad Adjacency recall &  0.7 & $(0.47,0.87)$ &  0.25 & $(0.067,0.467)$ & 0.001 \\ 
 \quad Orientation precision &  0.9 & $(0.5,1.0)$ &  0.52 & $(0,1)$ & 0.273 \\ 
 \quad Orientation recall &  0.5 & $(0.00,0.91)$ &  0.36 & $(0,1)$ & 0.316 \\ 
 \quad Proportion recovered v-structures &  0.3 & $(0.0,0.8)$ &  0.01 & $(0.00,0.14)$ & 0.106 \\ 
 \quad SID (lower bound) & 29.3 & $( 7,55)$ & 51.01 & $(29,74)$ & 0.072 \\ 
 \quad SID (upper bound) & 51.5 & $(22,81)$ & 58.43 & $(36,81)$ & 0.350 \\ 
   \hline
\end{tabular}
\caption*{\smaller Notes: CI denotes a 95\% confidence interval based on the empirical distribution. The $p$-values corresponds to one-sided tests.}
\end{table*}

We showcase the proposed simulation-based negative control procedure by evaluating the performance of the PC algorithm \citep{spirtes1991}. We construct a small toy simulation study considering the task of learning 10-node DAGs (or more specifically, CPDAGs corresponding to their Markov equivalence classes) from linear Gaussian data generated according to the DAGs. This is a scenario where the PC algorithm is sound and complete in the large sample limit, so in principle we should expect good performance. However, it is well-known that PC struggles on finite data when the true data generating mechanism is dense, because the algorithm is biased towards sparse graphs (see e.g., \citep{petersen2023sldisco}). We therefore consider a moderate sample size of $n = 400$ observations, and two settings for how dense the true DAGs are: A dense case with $\mtrue = 30$ edges and a sparse case with $\mtrue = 15$. We expect that PC performs better than negative controls in the sparse case, but not in the dense case. 

To evaluate PC, we consider five different metrics that make use of orientation information (and are hence beyond the scope of Sections \ref{sec.distrresults}-\ref{subsec.testfit}): Structural Hamming distance, orientation precision, orientation recall, proportion recovered v-structures, and structural intervention distance (SID) lower and upper bounds\footnote{The SID can only be reported as bounds as we are comparing a true DAG with an estimated CPDAG, see details in Supplementary Materials \ref{supl.pcdetails}.}. For comparability with the previous sections, we also report two adjacency metrics considered above; namely adjacency precision and recall. We provide definitions of the metrics and additional details about the simulation study in Supplementary Materials \ref{supl.pcdetails}. 

Table \ref{tab.pcsim} presents the results. In the dense case, we find that none of the metrics have sufficient discriminatory power to distinguish between PC and the negative control (testing at e.g., a 5\% significance level). For this case, PC estimated graphs with numbers of edges ranging from 7 to 19 with a mean of $13.3$, which is severely biased towards sparsity, as expected. 

For the sparse case, the number of edges estimated by PC ranges from 7 to 16 with a mean of $11.5$, i.e., a better match with $\mtrue$. We find that some metrics show significant differences between PC and the negative control, while others do not: SHD, and adjacency precision and recall are significantly better for PC than the negative controls (at a 5\% level), while the others are not. While the mean values for PC and negative controls are generally quite far apart, the 95\% confidence intervals reveal very broad distributions for orientation precision, orientation recall, proportion recovered v-structures and both SID bounds. Hence, there is a lot of variability in these metrics --- both for PC and negative controls --- and they are thus perhaps not very useful for evaluating this case. Another takeaway from this application is that reporting only mean values of metrics is not advisable; ideally, some description of the distribution (e.g., confidence intervals) should be included. 

Overall, we find that including negative controls can thus also be used to provide insights into the level of informativeness of a specific metric in scenarios where we have an established consensus of whether a certain causal discovery procedure "works well" or not.

\subsection{Application: Structural Hamming Distances on the Sachs Data}
\label{subsec.sachs}

\begin{table}[ht]
\centering
\caption{Structural Hamming Distances for the Sachs Data.}\label{tab.sachs}
\begin{tabular}{lcccc}
  \hline
 & \multicolumn{2}{c}{\textbf{Observed}} & \multicolumn{2}{c}{\textbf{Negative control}} \\
 \textbf{Algorithm} & SHD & $\mest$ & Mean SHD & $p$ \\ 
  \hline
PC & 23 & 24 & 31.54 & 0.001 \\ 
  NOTEARS & 22 & 16 & 27.10 & 0.050 \\ 
  LiNGAM & 30 & 33 & 34.43 & 0.083 \\ 
  GES & 30 & 30 & 34.24 & 0.114 \\ 
  BOSS & 35 & 32 & 35.24 & 0.510 \\    \hline
\end{tabular}
\end{table}

In this application, we consider the Sachs dataset \citep{sachs2005}, which is often used to evaluate causal discovery algorithms. The ground truth DAG for the Sachs dataset has 11 nodes and $\mtrue = 20$ edges\footnote{Note that there also exists an alternative version with only 17 edges.}. 

We evaluate the performance of five causal discovery procedures, namely PC \citep{spirtes1991}, GES \citep{chickering2002}, LiNGAM \citep{shimizu2006}, NOTEARS \citep{zheng2018} and BOSS \citep{andrews2023}. We apply each of these algorithms to the Sachs dataset and compute their SHD and estimated number of edges, $\mest$. Based on $\mest$, we simulate 1000 negative controls separately for each algorithm, and report the mean SHD over the negative controls. For algorithms that return a DAG (LiNGAM and NOTEARS), we simulate negative control DAGs, and for algorithms that return a CPDAG (PC, GES, and BOSS), we simulate negative control CPDAGs. In all cases we compare to the ground truth DAG by computing a one-sided $p$-value testing how often the discovery algortihm performs at least as well as the negative control (according to SHD). Additional details about the application are provided in Supplementary Materials \ref{supl.Sachsdetails}. 

Table \ref{tab.sachs} summarizes the results. We see that while the smallest SHD value is obtained by NOTEARS, PC produces the smallest $p$-value ($p = 0.001$) and hence is the furthest removed from random guessing, seconded by NOTEARS ($p = 0.050$) and LiNGAM ($p = 0.083$). In the other end of the spectrum, we find that GES, and especially BOSS, are not significantly different from random guessing testing at e.g., a 10\% significance level. 

This application illustrates that it is not very meaningful to compare and interpret differences in SHD without taking into account the number of edges placed. A lower SHD does not necessarily mean that an algorithm is further removed from random guessing; it may just reflect a more preferable level of sparsity in the estimated graph, which can be an artefact of the chosen metric. For example for SHD on the Sachs dataset, we can obtain $\text{SHD} = \mtrue = 20$ --- i.e., a value that outperforms all considered algorithms --- simply by "estimating" the empty graph. Hence, a more meaningful ranking may come about by considering the size of the negative control $p$-values. 

\section{Discussion}
\label{sec.disc}

The results presented here were developed with the aim of evaluating algorithms in an artificial "lab" setting where we have access to a known ground truth. Subsequent to such evaluations, the algorithms should of course also be tested in practice in real data applications without a known ground truth graph. How to asses performance in such scenarios is fundamentally difficult, as causal discovery is a unsupervised problem. One approach for validating causal discovery on real world data is to compare graphs (or resulting effect estimates) found by causal discovery algorithms with expert-made graphs based on theory or existing literature \citep{petersen2023,gururaghavendran2024}. However, such comparisons are of course only feasible when a comprehensive body of knowledge about the considered variables already exists, and even in this case, it is highly time consuming to construct expert graphs. A more broadly applicable evaluation strategy has been proposed by \cite{eulig2024}. They compare conditional independence fit of a candidate DAG (given by experts or estimated using causal discovery) with a baseline obtained by randomly permuting nodes. By doing so, they construct a statistical test of fit with a similar interpretation to the test proposed in Section \ref{subsec.testfit}, although focusing on conditional independencies implied by the DAG rather than edge presence directly. However, as the authors note, such a test is not readily applicable for validating causal discovery fit, as most algorithms use conditional independence information for estimating the graph, and hence a subsequent test based on the same information would result in overfitting. This could however be resolved by applying data splitting if the sample size renders such an approach feasible.

The distributional results for adjacency metrics presented in Section \ref{sec.distrresults} are conditional on three quantities: The maximal number of edges in the DAG ($\mmax$), the number of edges in the true DAG ($\mtrue$) and the number of edges in the estimated DAG ($\mest$). Clearly, conditioning on the first two is completely uncontroversial, but conditioning on $\mest$ may be debated. Our motivation for doing so is as follows: Many causal discovery algorithms --- e.g., PC \citep{spirtes1991}, FCI \citep{spirtes2000}, GES \citep{chickering2002}, and GRaSP \citep{lam2022} --- require choosing a tuning parameter, which will in practice directly control the number of outputted edges (e.g., test significance level in constraint-based algorithms or score penalties in score-based algorithms). Although we do not generally have a characterization of the relationship between the tuning parameter values and the resulting value of $\mest$, the relationship is generally deterministic in a single causal discovery application. The way causal discovery algorithms are mostly applied, the tuning parameter is \textit{not} chosen in a data-driven manner, but rather set at somewhat arbitrary "standard" values.  Alternatively, in some instances, it may be chosen based on external background knowledge \citep{petersen2023}. In either case, the number of edges in the estimated graph is \textit{de facto} chosen a priori, in which case we can meaningfully condition on it. 

Some work has been proposed for data-driven tuning of causal discovery algorithms \citep{biza2020}. If such methods are applied, $\mest$ will generally be estimated from data. In this case, we lose the distributional results for the adjacency metrics presented in Section \ref{sec.distrresults}, as the number of true positives will no longer be hypergeometrically distributed. But we can still use the proposed simulation-based pipeline from Section \ref{sec.simbasedNC} to obtain a negative control for such algorithms.

Another reason for preferring to condition on $\mest$ is related to interpretability. By only considering one value of $\mest$, we compare our causal discovery procedure with a well-specified negative control condition, namely placing the same number of edges at random. This argument also has implications for how we ought to compare two different causal discovery algorithms; to increase interpretability, we advise to either tune the algorithms to estimate the same number of edges and then compare their outputs, or follow the strategy from Section \ref{subsec.sachs} and compare negative control $p$-values. Otherwise, we may be comparing sparse outputs with dense ones without accounting for the different difficulties in estimating sparse and dense graphs, and as we have seen above, such a comparison may not be meaningful, and will definitely be difficult to interpret. Tuning algorithms to produce equally dense outputs has an additional benefit by removing the (difficult to interpret) tuning parameters from the evaluation equation altogether, and replacing them with the simpler notion of outputted graph density. 

However, we want to stress that it can be problematic to consider only a single density in a simulation-based evaluation study if the algorithm being evaluated is able to thereby learn the (unique) intended density \citep{petersen2023sldisco}. This evaluation design flaw has been present for several supervised discovery algorithms \citep{li2020,xu2021,yu2019}, and could harm transportability greatly. We propose that a range of densities should therefore always be considered when conducting simulation-based evaluation studies of discovery algorithms that may learn the density directly from training data. But to ease interpretation, the results should ideally be presented stratified according to density.

We have considered a range of different metrics that may be used to evaluate causal discovery methods, but the list is clearly not exhaustive. We have focused on edgewise and structural evaluations, as these are most commonly reported \citep{gentzel2019}, but we advice that all evaluation studies should include a critical consideration of what metrics are relevant for the specific intended use case. For example, if an intended usecase is mostly focused on causal information flow, and not whether effects are direct or indirect, it is natural to consider a metric that explicitly counts preserved ancestral information \citep{bang2024}. Or, if one is interested in using a causal discovery estimate for subsequent effect estimation and inference, one should include metrics on the intervention distribution, possibly targeting a specific causal estimand of interest \citep{gentzel2019}.

The work presented here has focused on Erdős-Rényi type graphs. This assumption is important for the distributional results in Sections \ref{sec.distrresults} - \ref{subsec.testfit}, as the hypergeometric distribution requires random draws. Non-central versions of the hypergeometric distribution allows for biased draws of edges, but we do not believe this is very useful for describing causal graphs: It would allow certain edges to be more likely to be present than others, but would still not consider graph properties beyond singular edges and hence not be appropriate for describing for example graphs that exhibit clustering. However, if a specific evaluation study wants to target such graphs, the simulation-based method proposed in Section \ref{sec.simbasedNC} can straightforwardly be applied, simply by simulating random graphs from the intended target graph type, see e.g., \citep{albieri2014}.  

As mentioned in Section \ref{subsec.testfit}, the exact distributional results for adjacency metrics will by definition result in conservative statistical inference, i.e. conservative control of type I error in statistical tests and overly wide confidence intervals. Due to the discrete nature of the hypergeometric distribution, this is especially pronounced when $\mmax$ is small, i.e. when there are only few nodes. However, we argue that the considered null hypothesis is very crude --- assuming completely random replacement of $\mest$ edges --- and hence we do not consider conservative inference to be very problematic. Informally, we would ideally like to perform markedly better than random guessing, not just borderline significantly so!

In conclusion, we believe the results and examples provided here showcase that we need to acknowledge that causal discovery is not just another machine learning problem. Estimating a high-dimensional object such as a graph is difficult, and evaluating how well one did is equally challenging. If we do not take into account the most fundamental property of the graphs we simulate for evaluation --- their densities --- we are not producing useful results that will be likely to generalize to new data with other graph densities. We believe that the use of negative controls will be a useful next step in the direction of more transparent and interpretable evaluations. We of course all hope to do better than random guessing, so let us make it easy to see when we do - and when we do not.

\begin{acknowledgements} 
	The author thanks Vanessa Didelez for her insightful feedback on this work. 
\end{acknowledgements}

\bibliography{negcontrol}

\newpage

\onecolumn

\title{Are You Doing Better Than Random Guessing? A Call for Using Negative Controls When Evaluating Causal Discovery Algorithms\\(Supplementary Material)}
\maketitle

\appendix

\section{Random Urn Experiment Motivation for Distributional Result}
\label{suppl.randomurn}

Consider Table \ref{tab.genericconf} and assume that all edges both in $G$ and $\hat{G}$ were placed uniformly at random. Then, given the number of true ($\mtrue$), estimated ($\mest$) and maximum total ($\mmax$) edges, the number of true positives can be seen as the result of a simple random urn experiment with two colors of balls, say, blue and white: White balls correspond to adjacencies included in $G$, and blue balls are adjacencies not in $G$. A random causal discovery procedure will then metaphorically draw "balls" (i.e., edges) randomly without replacement, and some will be true positives (white), while others will be false positives (blue). Since the number of white balls ($\mtrue$), the number of draws ($\mest$) and the total number of balls ($\mmax$) are all known a priori, the number of drawn white balls (true positive adjacencies) will by definition follow a hypergeometric distribution: $\TP  \cond \mmax, \mtrue, \mest \sim \text{HyperGeom}(\mmax, \mtrue, \mest)$.

\section{Computations for Table \ref{tab.adjmetrics}}
\label{suppl.adjmetrics}

Below, we let $q_{(k, \mmax, \mtrue, \mest)}$ be the $k$th quantile from $\text{HyperGeom}(\mmax, \mtrue, \mest)$. We also note, and use repeatedly below, that $\TP \cond (\mmax, \mtrue, \mest) \sim \text{HyperGeom}(\mmax, \mtrue, \mest)$, and hence $\E(\TP \cond \mmax, \mtrue, \mest) = \frac{\mest \cdot \mtrue}{\mmax}$. 

We will compute the expectation and quantiles for each of these four adjacency metrics: Recall, F1, negative predictive value (NPV), and specificity. 

\paragraph{Recall:}
We write recall as a function of $\TP, \mmax, \mtrue$ and $\mest$:
\begin{align*}
\text{recall} &= \frac{\TP}{\TP + \FN} \\
&=  \frac{\TP}{\TP + \mtrue - \TP} \\
&= \frac{\TP}{\mtrue}.
\end{align*}
Since this is a linear function of $\TP$, the conditional expectation given ($\mmax, \mtrue$, $\mest$) is
\begin{align*}
\E(\text{recall} \cond \mmax, \mtrue, \mest) &= \frac{1}{\mtrue} \E(\TP \cond \mmax, \mtrue, \mest) \\
&= \frac{1}{\mtrue}  \mest \frac{\mtrue}{\mmax} \\
&= \frac{\mest}{\mmax}
\end{align*}
and the $k$th quantile of the recall distribution, conditional on $(\mmax, \mtrue, \mest)$, is given by
\begin{equation*}
\frac{1}{\mtrue} q_{(k,\mmax, \mtrue, \mest)}.
\end{equation*}

\paragraph{F1:}
We write the F1 score as a function of $\TP, \mmax, \mtrue$ and $\mest$:
\begin{align*}
\text{F1} &=  \frac{2 \cdot \TP}{2 \cdot \TP + \FP + \FN} \\
&= \frac{2 \cdot \TP}{2 \cdot \TP + \mest - \TP + \mtrue - \TP} \\
&= \frac{ 2 \cdot \TP}{\mest + \mtrue}.
\end{align*}
Since this is a linear function of $\TP$, the conditional expectation given ($\mmax, \mtrue$, $\mest$) is
\begin{align*}
\E(\text{F1} \cond \mmax, \mtrue, \mest) &= \frac{ 2 \cdot \E(\TP \cond \mmax, \mtrue, \mest)}{\mest + \mtrue} \\
&= \frac{2 \cdot \frac{\mest \cdot \mtrue}{\mmax}}{\mest + \mtrue} \\
&= \frac{2 \cdot \mest  \cdot \mtrue}{\mmax \cdot (\mest + \mtrue)}
\end{align*}
and the $k$th quantile of the F1 distribution, conditional on $(\mmax, \mtrue, \mest)$, is given by
\begin{equation*}
\frac{ 2 \cdot q_{(k,\mmax, \mtrue, \mest)}}{\mest + \mtrue}.
\end{equation*}

\paragraph{NPV:}
We write the negative predictive value (NPV) as a function of $\TP, \mmax, \mtrue$ and $\mest$:
\begin{align*}
\text{NPV} &= \frac{\TN}{\TN + \FN} \\
&= \frac{\mmax - \mest - \mtrue + TP}{\mmax - \mest - \mtrue + \TP + \FN} \\
&= \frac{\mmax - \mest - \mtrue + TP}{\mmax - \mest}.
\end{align*}
Since this is a linear function of $\TP$, the conditional expectation given ($\mmax, \mtrue$, $\mest$) is
\begin{align*}
\E(\text{NPV} \cond \mmax, \mtrue, \mest) &= \frac{\mmax - \mest - \mtrue + \E(\TP \cond \mmax, \mtrue, \mest)}{\mmax - \mest}\\
&= \frac{\mmax - \mest - \mtrue + \frac{\mest \cdot \mtrue}{\mmax}}{\mmax - \mest} \\
&= 1 - \frac{\mtrue}{\mmax} \\
&= 1 - \E(\text{precision} \cond \mmax, \mtrue, \mest)
\end{align*}
and the $k$th quantile of the NPV distribution, conditional on $(\mmax, \mtrue, \mest)$, is given by
\begin{equation*}
\frac{\mmax - \mest - \mtrue + q_{(k,\mmax, \mtrue, \mest)}}{\mmax - \mest}.
\end{equation*}

\paragraph{Specificity:}
We write specificity as a function of $\TP, \mmax, \mtrue$ and $\mest$:
\begin{align*}
\text{specificity} &=  \frac{\TN}{\TN + \FP}\\
&= \frac{\mmax - \mest - \mtrue + \TP}{\mmax - \mest - \mtrue + \TP + \FP} \\
&= \frac{\mmax - \mest - \mtrue + \TP}{\mmax - \mtrue}.
\end{align*}
Since this is a linear function of $\TP$, the conditional expectation given ($\mmax, \mtrue$, $\mest$) is
\begin{align*}
\E(\text{specificity} \cond \mmax, \mtrue, \mest) &= \frac{\mmax - \mest - \mtrue + \E(\TP \cond \mmax, \mtrue, \mest)}{\mmax - \mtrue}\\
&= \frac{\mmax - \mest - \mtrue + \frac{\mest \cdot \mtrue}{\mmax}}{\mmax - \mtrue}\\
&= 1 - \frac{\mest}{\mmax} \\
&= 1 - \E(\text{recall} \cond \mmax, \mtrue, \mest)
\end{align*}
and the $k$th quantile of the specificity distribution, conditional on $(\mmax, \mtrue, \mest)$, is given by
\begin{equation*}
\frac{\mmax - \mest - \mtrue + q_{(k,\mmax, \mtrue, \mest)}}{\mmax - \mtrue}.
\end{equation*}

\section{Details About PC Algorithm Simulation Study}
\label{supl.pcdetails}

Let $G$ be the true graph and $\hat{G}$ be an estimate. We consider the following metrics: 
\begin{description}
\item[Structural Hamming distance:] The structural Hamming distance (SHD) counts the number of edge reversals, removals and additions needed in order to transform $\hat{G}$ into $G$ \citep{tsamardinos2006}.
\item[Adjacency precision and recall:] Precision and recall computed from the adjacency/skeleton confusion matrix, as described in Section \ref{sec.precrecall}. 
\item[Orientation precision and recall:] Precision and recall computed from a conditional orientation confusion matrix. The conditional orientation confusion matrix is contructed as follows: For all edges that are both in $G$ and $\hat{G}$, each edge endpoint is classified as: 
\begin{itemize}
\item True positive if there is an arrowhead both in $G$ and $\hat{G}$. 
\item True negative if there is a tail both in $G$ and $\hat{G}$.
\item False positive if there is an arrowhead in $\hat{G}$, but a tail in $G$. 
\item False negative if there is a tail in $\hat{G}$, but an arrowhead in $G$. 
\end{itemize}
\item[Proportion recovered v-structures:] v-structures are node triples with the structure $A \to B \leftarrow C$ where $A$ and $C$ are non-adjacent. This metric counts how many such structures are correctly recovered by $\hat{G}$, divided by the total number of v-structures in $G$. If there are no v-structures in $G$, the value is set to 1 (interpreted as all structures being recovered). 
\item[Structural intervention distance:] The structural intervention distance (SID) counts the number of node pairs $(X_i, X_j)$ for which $\hat{G}$ does not provide a valid adjustment set for the total causal effect of $X_i$ on $X_j$ (assuming $G$ is the truth) \citep{peters2015}. This property is only well-defined for DAG-DAG comparisons: If $\hat{G}$ is a CPDAG, it may have undirected edges and hence the total causal effects may not be identifed. In this case, one instead computes the SID for each DAG in the equivalence class specified by the CPDAG and reports the minimum and maximum values as bounds. For computing SIDs, we use the \texttt{SID} R package \citep{peters2023} with default settings\footnote{Note that some cases result in warnings due to the estimated graph not being a proper CPDAG, or having large connected components. In both cases, SID is computed based on local expansions of the graphs.}. 
\end{description}

We now provide a step-by-step example of how the negative controls are computed and used, following the template from Section \ref{sec.simbasedNC}. We focus on a single metric (SHD) and setting (dense). We proceed as follows: 
\begin{description}
\item[1. Standard simulation study:] We draw 1000 random Erdős-Rényi type DAGs over 10 nodes, each with $\mtrue = 30$ edges. From each DAG, we simulate 400 independent Gaussian observations with randomly drawn regression parameters and error variances\footnote{We use default options for regression parameters and error variances from the \texttt{simGausFromDAG()} function in the \texttt{causalDisco} R package \citep{petersen2022}.}. For each dataset, we apply the PC algorithm using a test for vanishing partial correlations with significance level $\alpha = 0.05$.
\begin{enumerate}
\item[(a)] We compare each of the 1000 true DAGs with the corresponding estimated CPDAG provided by the PC algorithm by computing their SHDs. We find a mean SHD of $27.33$ with a 95\% confidence interval of $(21;33)$ (based on the empirical quantiles). 
\item[(b)] We store the true DAGs as well as the distribution of the estimated number of edges across the 1000 applications of the PC algorithm. The number of estimated edges from PC range from 7 to 19 with a mean of $13.3$.
\end{enumerate}
\item[2. Negative control simulation:] We draw  1000 random CPDAGs over 10 nodes with number of edges independently sampled from the $\mest$ distribution from Step 1 (b). 
\item[3. Negative control evaluation:] For each of the 1000 negative controls, we compare with a true DAG from Step 1 by computing the SHD. We find a mean SHD of $30.23$ (95\% CI: $26;36$).
\item[4. Comparison:] We conduct pairwise comparisons of SHD values obtained by PC and negative controls. We find
\begin{equation*}
p = \frac{1}{1000}\sum_{i = 1}^{1000} \mathbf{1} \left(\text{SHD}(G_\text{true}^i, \hat{G}_\text{PC}^i)  \geq \text{SHD}(G_\text{true}^i, \hat{G}_\text{NC}^i)\right) = 0.202
\end{equation*}
and hence conclude that PC is not significantly different from the negative controls.
\end{description}

\section{Details About Sachs Data Application}
\label{supl.Sachsdetails}

For PC, LiNGAM, GES and BOSS, we apply the algorithms in Tetrad version 7.6.7 \citep{ramsey2018} with default settings. For NOTEARS, we report SHD and number of edges as provided in \citep{zheng2018}.  

The negative control DAGs are simulated as Erdős-Rényi type DAGs. The negative control CPDAGs are constructed by first simulating an Erdős-Rényi type DAG, and secondly constructing the CPDAG corresponding to its Markov equivalence class. 

We use the 20 edge "truth" version of the Sachs dataset because that is also what was used to evaluate the NOTEARS algorithm in \citep{zheng2018}. We obtain both the ground truth graph and the Sachs dataset from this repository: \url{https://github.com/cmu-phil/example-causal-datasets}.

\end{document}